 \documentclass[final,3p,times,twocolumn]{elsarticle}

\usepackage{caption}
\usepackage{subcaption}
\usepackage{amssymb}
\usepackage{amsmath}
\usepackage{svg}
\usepackage{blindtext}
\usepackage{lineno}
\usepackage{hyperref}
\usepackage{pdflscape}
\usepackage{comment}
\usepackage{booktabs}
\usepackage{graphicx}
\usepackage{epstopdf}
\hypersetup{
    colorlinks=true,
    linkcolor=blue,
    filecolor=magenta,      
    urlcolor=cyan,
}

\journal{Control Engineering Practice}

\begin{document}

\begin{frontmatter}


\title{Model-Free Control approach for pH Regulation in Thin-Layer Photobioreactors}

\author[1]{José González-Hernández*}
\author[2]{Ainoa Morillas-España}
\author[1]{José Luis Guzmán}
\author[1]{José Carlos Moreno}
\author[3]{Alain Vande Wouwer}

\address[1]{University of Almería, Department of Informatics, CIESOL, ceiA3, La Cañada de San Urbano s/n, Almería, Spain. }
\address[2]{University of Almería, Chemical Engineering, CIESOL, ceiA3, La Cañada de San Urbano s/n, Almería, Spain. }
\address[3]{Systems, Estimation, Control and Optimization (SECO), University of Mons, B-7000, Mons, Belgium.\\ *Corresponding author: José González-Hernández (j.gonzalez@ual.es)}

\begin{abstract}
Thin-layer photobioreactors (TLRs) exhibit fast hydrodynamic and thermal dynamics,
strong nonlinear photosynthetic responses and significant time-variability due to
irradiance fluctuations and biomass growth. These characteristics challenge
conventional model-based control strategies, whose tuning degrades under rapidly
changing operating conditions. This work presents the experimental implementation of a model-free control approach, Extremum Seeking Control (ESC), for performance optimization in a semi-industrial thin-layer photobioreactor. Unlike previous studies in raceway ponds, the reduced hydraulic inertia of TLR systems enables the adaptation of this control strategy to accelerate convergence while preserving gradient estimation accuracy. The proposed approach is experimentally compared against classical on-off control and ESC configurations with and without feedforward compensation of solar irradiance. Beyond control performance metrics, biological indicators such as biomass concentration and productivity are evaluated to assess the impact on process efficiency. Results show that the proposed ESC strategy reduced cumulative CO$_2$ consumption by approximately 39\% and decreased the accumulated pH tracking error by more than 60\% compared with conventional on-off control, while biomass- and irradiance-normalised indicators confirmed a more efficient use of injected carbon. These results demonstrate that high-frequency ESC can improve regulation performance and carbon utilisation efficiency in fast photobioreactor systems, highlighting its suitability for thin-layer cultivation under outdoor conditions.
\end{abstract}

\begin{keyword}
real-time optimization \sep process control \sep extremum seeking control \sep microalgae \sep thin-layer reactor

\end{keyword}

\end{frontmatter}


\section{Introduction}

Microalgae are photosynthetic microorganisms that have attracted renewed interest in recent years because of their potential to address pressing environmental and resource challenges \citep{interes_Gabriel,produccion,Scenedesmus_wastewater}. Their ability to capture carbon dioxide (CO$_2$) and convert it into biomass contributes to climate-change mitigation through net greenhouse-gas removal, while their rapid growth in diverse environments supports applications ranging from wastewater treatment to sustainable biomass production.

Microalgae are typically cultivated in photobioreactors, which are broadly classified as open or closed systems. Open systems, such as raceway ponds, are widely used in industrial cultivation due to low cost and simple operation, especially when nutrient levels are sufficient and contamination control is less critical \citep{raceway_acien,outdoor_raceway}. Closed systems, including tubular and flat-panel designs, facilitate tighter control of cultivation conditions and are prevalent when biomass quality or product specificity is required.

Among open systems, raceway photobioreactors dominate industrial-scale cultivation. However, their large characteristic time constants, slow recirculation and strong coupling with weather conditions complicate regulatory control, particularly for pH, a variable strongly linked to photosynthetic performance and microalgal health \citep{ABACO,GuzACC2025}. The pH is usually regulated by injecting CO$_2$, which acidifies the culture and supplies inorganic carbon; hence, effective pH control has been the focus of much prior work \citep{rejection,Wiener,daytime}.

A significant body of research has explored model-based and adaptive control strategies for pH regulation in raceway systems, including PI loops with periodic retuning \citep{Caparroz2025a,Caparroz2025b}, data-driven surrogates wrapped around classical controllers \citep{Caparroz2024}, hybrid MRAC schemes, and learning-based model predictive approaches \citep{Pataro2023,3DOF_Otalora}. While these strategies accommodate slow seasonal changes and slowly varying dynamics, they still depend on some form of model identification or offline parameter adjustment, which becomes expensive and fragile under large diurnal forcing and evolving culture conditions.

In contrast, control strategies for thin-layer photobioreactors (TLRs) have received comparatively little attention in the control literature despite growing bioprocess interest. TLRs operate with a much reduced culture depth, resulting in enhanced light penetration and intensified hydrodynamics, consistent with observations in shallow photobioreactor configurations \citep{Ugwu2008}. However, these features also lead to increased sensitivity to environmental disturbances and faster, highly nonlinear metabolic responses \citep{Bernard2011}.

Recent experimental studies on microalgae-based wastewater treatment systems have confirmed their high productivity and operational potential \citep{Christenson2011,Park2011}. In thin-layer configurations, these advantages are further enhanced due to improved light availability and faster hydrodynamics, leading to distinct dynamic behavior compared to raceway reactors. In particular, comparative analyses highlight substantially faster dynamics and stronger irradiance-metabolism coupling in TLRs, reinforcing the need for dedicated control strategies \citep{MJ_thinlayer}.

Despite these advances, most existing works primarily focus on process characterization, productivity assessment, or basic control approaches such as PI and on–off strategies. This lack of advanced control methodologies is particularly critical in TLR systems, where dominant time scales are significantly shorter and conventional controllers may fail to react within relevant biological time scales. Notably, model-free and optimization-based control strategies remain largely unexplored in this context, despite their potential to handle fast, nonlinear, and poorly modelled dynamics inherent to TLR systems.

The relative scarcity of control research addressing TLRs, and particularly strategies capable of handling their rapidly varying and uncertain dynamics, motivates the present work. Thin layers exhibit dominant time scales that are much shorter than those of raceway systems, due to reduced hydraulic inertia and more intense coupling between irradiance and metabolic activity. Such properties challenge conventional PI or model-based controllers, but also open the door for model-free approaches that can exploit real-time measurements without relying on an explicit process model.

Extremum seeking control (ESC) is one such model-free strategy: by injecting a small periodic perturbation and demodulating the measured performance signal, the controller estimates the local gradient online and steers the operating point toward the value that maximizes a chosen objective, all without an explicit model and while adapting to time-varying dynamics \citep{Ariyur2003,Krstic2000,Tan2006}. ESC has been applied successfully in various fields, including power systems, combustion, and fluid dynamics \citep{Dewasme2020}, and has been demonstrated in photobioreactor systems \citep{feudjio2021,feudjio2021dual,lara2025cheap}. In particular, recent experimental work has validated ESC for pH regulation in semi-industrial microalgae raceway reactors, highlighting its capability for real-time optimization under slow and time-varying dynamics \citep{GonzalezHernandez2026}. However, these systems are characterized by relatively slow hydrodynamics and large time constants. To the best of our knowledge, the application of ESC to thin-layer photobioreactors, which exhibit significantly faster dynamics and stronger coupling with irradiance, has not been explored.

In this work, we investigate the deployment of a ESC architecture for model-free pH control in a thin-layer photobioreactor. By leveraging the reduced dominant time constant of the system, we explore controller configurations that enable rapid gradient estimation and robust tracking under strong environmental variability. Specifically, we compare ESC strategies with feedforward irradiance compensation against classical on-off pH control, and evaluate both control performance and biological outcomes such as biomass concentration and specific energy consumption per unit biomass.

This paper is structured as follows. Section~2 presents the characterization of thin-layer dynamics, the experimental reactor setup, and the high-frequency ESC design. Section~3 describes the comparative evaluation of the control strategies in terms of convergence performance and biological metrics. Section~4 discusses the implications of the results, emphasizing the suitability of model-free control in fast, time-varying photobioreactor systems. Section~5 concludes the paper with insights and future research directions.

\section{Materials and methods}
This section describes the experimental platform, the instrumentation and control architecture, and the methodological framework used for the implementation and evaluation of the proposed control strategy. First, the thin-layer photobioreactor and its operating conditions are presented. Then, the control system, the ESC formulation, the baseline on-off controller, and the biomass measurement procedure are described.

\subsection{Thin-layer photobioreactor}

\begin{figure}[ht]
    \centering
    \includegraphics[width=0.95\linewidth]{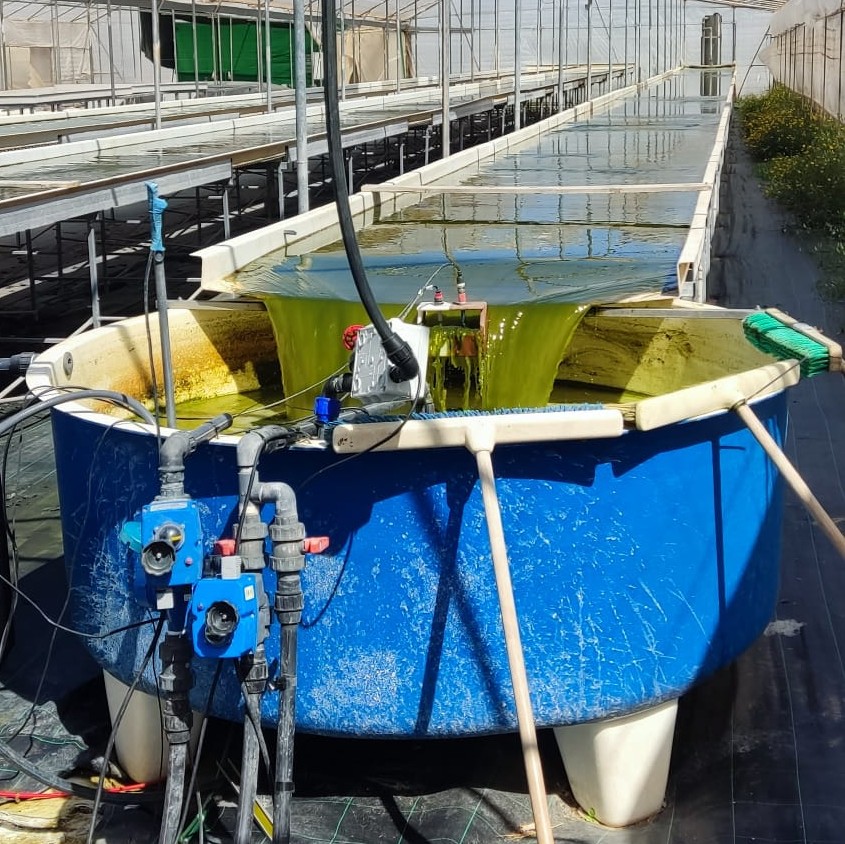}
    \caption{Thin-layer photobioreactor (63$\mathrm{m^2}$), located at the IFAPA facilities in Almería, Spain.}
    \label{fig:TLR}
\end{figure}

The experiments were conducted in an outdoor TLR located at the facilities of IFAPA in Almería, Spain. The reactor was operated with a \textit{Chlorella} culture under outdoor conditions. A schematic view of the system is provided in Fig.~\ref{fig:TLR}.

The reactor consists of a shallow open-channel section with a surface area of $63\,\mathrm{m^2}$ and a culture depth of $0.02\,\mathrm{m}$, resulting in a working volume of about $1.26\,\mathrm{m^3}$. The installation also includes a collection tank of $1.8\,\mathrm{m^3}$, where gas injection is performed, and an auxiliary column of $0.23\,\mathrm{m^3}$ located at the opposite end of the reactor.

The culture flows along the open channel into the collection tank, from which it is pumped to the auxiliary column. This column acts as a hydraulic buffer, ensuring continuous recirculation and stable feeding conditions at the channel inlet.

Due to its shallow configuration, the system exhibits high sensitivity to variations in biomass concentration and environmental conditions. In particular, the reactor response is strongly influenced by culture concentration, affecting both the forced response associated with $\mathrm{CO_2}$ injection and the natural response driven by photosynthetic activity, which depends on incident irradiance and light attenuation.

As a result, the system displays strongly time-varying and nonlinear behavior, making it suitable for the evaluation of model-free optimization and control strategies.

\subsection{Control system and implementation}

The reactor is managed by a Modicon M241 programmable logic controller, which integrates an OPC UA server for real-time data exchange. Through this interface, sensor measurements are acquired and actuator commands are issued.

The measured variables considered in this work are the pH value, obtained from a Crison 50 10 T probe, and the global solar irradiance, measured using a Kipp \& Zonen CM 6B pyranometer. The manipulated variable is the $\mathrm{CO_2}$ injection flow rate, regulated by a FESTO VEMD proportional valve equipped with an internal control loop that enables operation via a flow-rate setpoint.

All process data are transmitted via Ethernet to an industrial network, allowing access from supervisory systems and external controllers.

A desktop computer running MATLAB is used to implement the ESC algorithm. The control strategy is executed through a read-compute-write loop implemented in MATLAB. At each sampling instant, MATLAB queries the OPC UA server to retrieve the process measurements, computes the corresponding control action, and writes the updated $\mathrm{CO_2}$ flow setpoint back to the control system via the OPC UA server.

The control loop operates with a fixed sampling time of 10 seconds, selected as a compromise between capturing the fast process dynamics of the TLR and ensuring reliable communication and computation.

In this work, ESC is employed as a model-free optimization strategy to regulate the $\mathrm{CO_2}$ injection based solely on process measurements, without relying on an explicit model of the biological or physicochemical dynamics. The detailed formulation of the ESC algorithm is presented in the following section.

\subsection{Extremum-Seeking Control (ESC)}
\label{subsec:esc_classic}

ESC is a model-free optimization approach that allows a process to be automatically driven toward an optimal operating condition using only real-time measurements. Rather than relying on a detailed model of the photobioreactor, ESC introduces small changes in the control input and observes how a selected performance index responds. From this information, the controller estimates whether the operating point should be increased or decreased. This feature is especially useful in microalgae cultivation, where biological activity, light availability, and environmental disturbances make the system difficult to model accurately.

Thus, ESC is employed to perform real-time optimization of a scalar performance index using only process measurements, without relying on an explicit model of the system dynamics. In the following, the general idea of ESC is summarized \citep{Dewasme2020}, and in Section 3.3 is adapted for the specific case of this work focused on pH control.

Let \(y \in \mathbb{R}\) denote the measured process output. The control objective is defined through a cost function
\begin{equation}
J(t) = \Psi\!\big(y\big),
\end{equation}
where \(\Psi(\cdot)\) represents a mapping from the measured variables to a performance metric of interest.

In contrast to steady-state optimization approaches, ESC operates directly on the time-varying signal \(J\), which contains both slow process variations and oscillatory components induced by the control action.

\begin{figure}[h]
  \centering
  \includegraphics[width=0.95\linewidth]{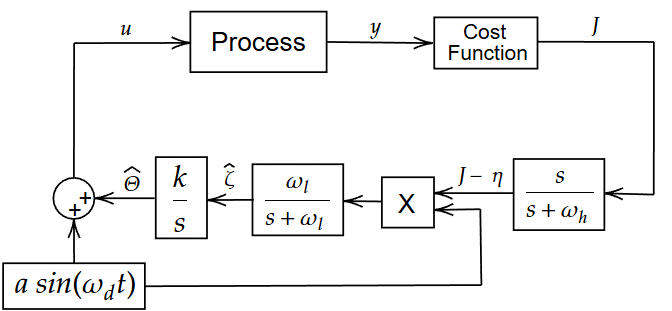}
  \caption{Classical modulation--demodulation ESC:a sinusoidal dither \(a\sin(\omega_d t)\) is added to the estimated operating point \(\hat{\theta}\) to generate the process input \(u\). The process output is mapped to a cost \(J\), high-pass filtered (washout \(s/(s+\omega_h)\)), multiplied by the reference dither (block X), low-pass filtered \(\big(\omega_l/(s+\omega_l)\big)\), and finally integrated with gain \(k/s\) to update the operating point.}
  \label{fig:esc_blockdiagram}
\end{figure}

A classical modulation--demodulation ESC scheme is adopted in this work to drive the system toward an extremum of \(J\). The method relies on injecting a small periodic perturbation into the manipulated variable and extracting gradient information from the measured response. The general structure of the controller is illustrated in Fig.~\ref{fig:esc_blockdiagram}.

For a single input, the ESC dynamics can be expressed as
\begin{align}
u &= \hat{\theta} + a\,\sin(\omega_d t), \\[2mm]
\dot{\hat{\theta}} &= k\,\hat{\zeta}, \\[2mm]
\dot{\hat{\zeta}} &= -\omega_l \hat{\zeta}
+ \omega_l (J - \eta)\, a\, \sin(\omega_d t), \\[2mm]
\dot{\eta} &= -\omega_h \eta + \omega_h J.
\end{align}

The control input \(u\) is obtained by superimposing a sinusoidal perturbation of amplitude \(a\) and frequency \(\omega_d\) onto the slowly varying estimate \(\hat{\theta}\). The measured cost signal \(J\) is processed through a high-pass filtering operation implemented via subtraction of its low-frequency component \(\eta\), which is generated by a first-order low-pass filter.

The resulting signal \((J - \eta)\) captures the variations induced by the injected perturbation and is multiplied by the reference sinusoid to perform demodulation. This operation provides an estimate of the local gradient direction. A subsequent low-pass filtering stage removes high-frequency components, yielding a smooth estimate \(\hat{\zeta}\), which is integrated to update the parameter \(\hat{\theta}\).

The main tuning parameters of the controller are the perturbation amplitude \(a\), the dither frequency \(\omega_d\), the signal-conditioning parameters, and the adaptation gain \(k\). Their practical selection is described later in the context of the experimental ESC implementation.

\subsection{Baseline control strategy}
\label{subsec:onoff_control}

For comparison purposes, a conventional on-off control strategy was implemented, as commonly used in industrial microalgae cultivation systems due to its simplicity and robustness under time-varying operating conditions.

A setpoint of 8.0 was defined, which lies within the pH range reported as suitable for \textit{Chlorella vulgaris} growth \citep{sakarika2016effect}, with a hysteresis band of $\pm 0.1$. Thus, CO$_2$ injection is activated when the pH exceeds 8.1 and stopped when it falls below 7.9, resulting in a hysteresis-based control law. The CO$_2$ flow rate during injection is fixed at $8\,\mathrm{L\,min^{-1}}$.

This control strategy provides a simple and robust regulation of pH, but does not explicitly optimize process performance. Instead, it maintains the system within an acceptable operating range, but with a potentially suboptimal CO$_2$ utilization and increased variability in the process response.


\subsection{Biomass concentration measurements}
\label{subsec:biomass}

Biomass concentration was periodically determined through dry weight measurements in order to obtain an independent estimation of culture growth and to enable the evaluation of process performance in terms of carbon utilization.

Biomass concentration ($C_b$, g~L$^{-1}$) was determined gravimetrically. Triplicate culture samples of 100~mL were collected and filtered through pre-weighed glass microfibre filters with a nominal pore size of 1~$\mu$m. Filters were subsequently dried in a convection oven at 80~$^\circ$C for 24~h until constant weight was reached. Biomass concentration was calculated as the difference between the dry filter weight and the initial tare weight, divided by the volume of culture filtered. These measurements allow quantifying significant and rapid variations in biomass concentration occurring during operation, which directly affects the process dynamics and system response. This aspect is particularly relevant in the present work, as it highlights the ability of the proposed control strategy to operate under strongly varying conditions without relying on an explicit process model, since the control action is based solely on real-time measurements.

Due to the offline nature of dry weight measurements, these data were not used for real-time control but exclusively for post-process analysis and performance assessment, enabling the definition of performance metrics that relate CO$_2$ utilization efficiency to biomass concentration.

\section{Results and Discussion}
This section presents the experimental results obtained during the characterization and closed-loop evaluation of the thin-layer photobioreactor. The analysis first focuses on the dynamic response of the system under sinusoidal CO$_2$ excitation, which motivates the ESC tuning. The baseline on-off performance is then evaluated, followed by the design, validation, and quantitative comparison of the proposed ESC strategy.

\subsection{System characterization}

The dynamic behavior of the thin-layer photobioreactor was experimentally characterized in order to inform the design and tuning of the ESC strategy. In particular, the relationship between CO$_2$ injection and pH response was analyzed under controlled excitation signals to identify the relevant time scales and frequency-dependent behavior of the system.

\begin{figure*}[h]
  \centering
  \includegraphics[width=0.99\linewidth]{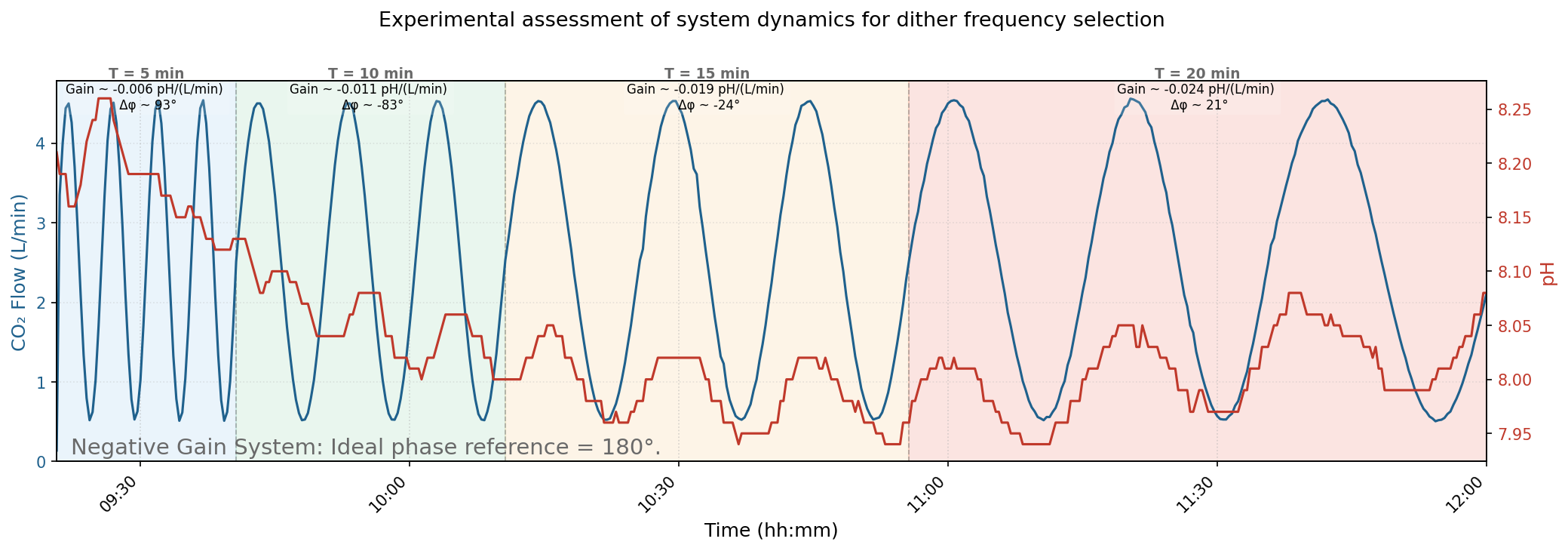}
  \caption{Experimental assessment of system dynamics under sinusoidal CO$_2$ excitation with different periods.}
  \label{fig:dither_dynamics}
\end{figure*}

A set of sinusoidal excitation experiments was conducted by modulating the CO$_2$ flow rate with different periods (5, 10, 15, and 20 minutes). Figure~\ref{fig:dither_dynamics} shows representative time series of the input (CO$_2$ flow) and the corresponding output (pH). These experiments allow a direct assessment of the system response in terms of amplitude attenuation and phase shift.

A clear frequency-dependent behavior was observed. At short periods (5 min), the pH response is strongly attenuated, indicating that the biological and physicochemical dynamics are too slow to follow rapid variations in CO$_2$ injection. This results in a very low effective gain and a significant phase lag, making the input--output relationship poorly correlated. As the excitation period increases (10 min), the system begins to respond more clearly, although attenuation and phase shift are still significant.

For intermediate periods (15 min), the system exhibits a substantially improved response, with increased output amplitude and reduced phase lag. This operating point provides a clear and consistent correlation between the excitation signal and the pH response, which is essential for reliable gradient estimation in ESC. In contrast, at longer periods (20 min), although the amplitude response remains high and the phase lag is further reduced, the system becomes increasingly affected by slow biological drift and non-stationary effects, which distort the input-output relationship over time.

The estimated gain and phase shift for each excitation period, as indicated in Figure~\ref{fig:dither_dynamics}, further support these observations. The system exhibits a negative gain, consistent with the expected inverse relationship between CO$_2$ injection and pH. Consequently, an ideal phase reference of 180$^\circ$ is required for correct gradient interpretation. At short periods, the large phase deviation from this reference compromises the effectiveness of the ESC scheme. At intermediate periods, particularly at 15 minutes, the phase shift approaches a more favorable range, enabling accurate gradient estimation while maintaining sufficient excitation of the system.

These results highlight the existence of a limited frequency band in which the system can be effectively excited without being dominated either by high-frequency attenuation or low-frequency drift. Based on this analysis, a dither period of 15 minutes was selected as a suitable compromise between sensitivity and robustness for subsequent ESC implementation. Although the analysis is based on representative experimental segments, the observed trends were consistent across repeated experiments.

Overall, the characterization confirms that the thin-layer reactor exhibits relatively fast but non-negligible dynamics, with a strong dependence on the excitation frequency. This behavior must be explicitly considered in the design of real-time optimization strategies such as ESC.

\subsection{Baseline performance (on-off control)}

\begin{figure*}[t]
    \centering
    \includegraphics[width=0.9\linewidth]{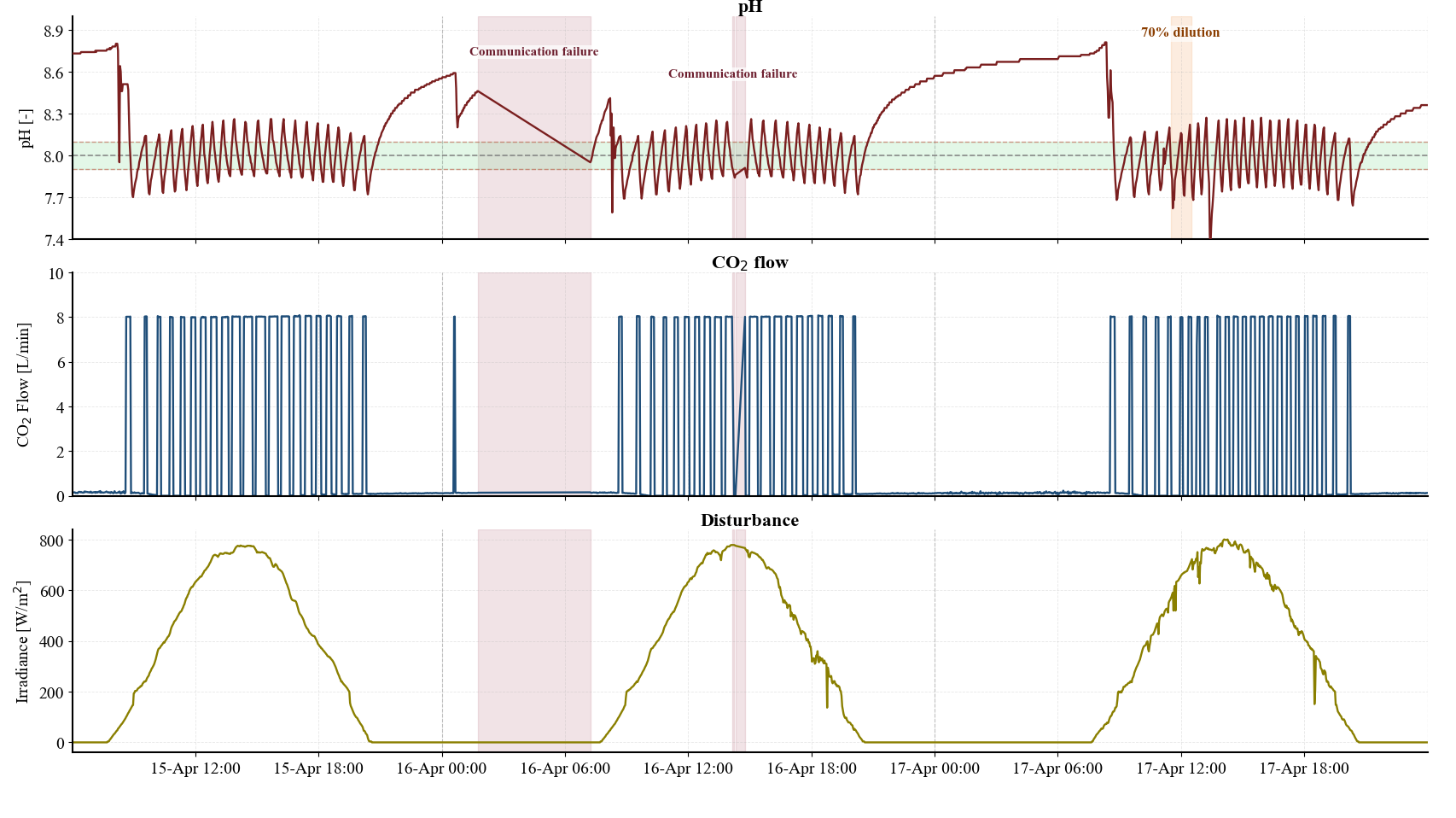}
    \caption{Three-day operation under on-off control showing pH (top), CO$_2$ flow (middle), and irradiance (bottom). Shaded regions indicate communication failures and dilution events.}
    \label{fig:onoff_response}
\end{figure*}

The performance of the baseline on-off control strategy was evaluated during three consecutive days of continuous operation. Figure~\ref{fig:onoff_response} illustrates the temporal evolution of pH, CO$_2$ injection, and irradiance.

As expected from discontinuous actuation, the controller induces persistent oscillations around the desired pH range. These oscillations arise from the cyclic switching between maximum CO$_2$ injection and complete shutdown, leading to a characteristic limit-cycle behavior. The amplitude and frequency of these oscillations are not constant but evolve over time, reflecting the strong coupling between control action and biological activity.

A key observation is the clear dependence of the system dynamics on irradiance. During daylight periods, increased photosynthetic activity accelerates carbon consumption, resulting in faster pH rises and higher switching frequency. In contrast, reductions in irradiance (e.g., sunset or transient shading) decrease metabolic activity, altering both oscillation patterns and CO$_2$ duty cycles. This confirms that the closed-loop response is inherently non-stationary.

Shaded regions in the figure highlight communication failures, during which CO$_2$ injection is interrupted. These events lead to noticeable pH drifts and delayed recovery once communication is restored, revealing limited disturbance rejection capability under actuator unavailability.

Moreover, the pulsed nature of the control action results in non-uniform carbon dosing. The system alternates between over-injection and inactivity, preventing operation near an optimal equilibrium. This effect becomes more evident after events such as dilution, where the controller responds aggressively but inefficiently, overshooting before returning to its oscillatory regime.

In summary, while the on-off controller provides a simple and robust mechanism for maintaining bounded operation, it inherently produces sustained oscillations, exhibits time-varying performance, and leads to inefficient CO$_2$ utilization. These structural limitations justify its use as a baseline for comparison with the proposed extremum seeking control strategy. This inefficiency is quantified in Section~\ref{sec:esc_performance} through cumulative CO$_2$ consumption metrics.

\subsection{ESC design and modifications}

An ESC strategy was implemented to enable real-time regulation based on pH measurements, without relying on an explicit process model. The objective of the controller is to steer the system towards operating conditions that minimize the deviation between the measured pH and the desired setpoint under varying culture conditions.

\subsubsection{Baseline ESC structure}

The implemented ESC scheme follows the classical perturbation-based approach presented in Section 2.3. A sinusoidal dither signal is superimposed on the manipulated variable (CO$_2$ flow rate), inducing small periodic variations in the system. The resulting pH response is then demodulated to estimate the local gradient of the objective function with respect to the input.

This gradient estimate is used to update the control input through an integral adaptation law, driving the system towards the extremum. Based on the system characterization presented in the previous section, a dither period of 15 minutes was selected as a compromise between sufficient system excitation and robustness to slow biological drift.

The remaining ESC parameters were selected experimentally to ensure stable convergence without inducing excessive pH oscillations. The perturbation amplitude was chosen to provide a measurable pH response while keeping the CO$_2$ flow variations within a range compatible with normal reactor operation. The adaptation gain was tuned conservatively to avoid aggressive changes in the operating point, particularly under rapid irradiance variations. 


\subsubsection{Limitations under non-stationary conditions}

In practice, the application of standard ESC to the photobioreactor revealed important limitations. The pH signal exhibits slow time-varying trends due to biological activity and environmental fluctuations, leading to non-stationary operating conditions. These low-frequency components interfere with the demodulation process and degrade the quality of the gradient estimation.

In conventional ESC implementations, a high-pass filter is typically used to remove low-frequency components prior to demodulation. However, in this system, the high-pass filtering approach proved insufficient. The presence of slow drift combined with relatively low dither frequencies resulted in partial overlap between the excitation signal and the underlying trend, leading to distorted gradient estimates and reduced convergence performance.

\begin{figure}[t]
    \centering
    \includegraphics[width=\linewidth]{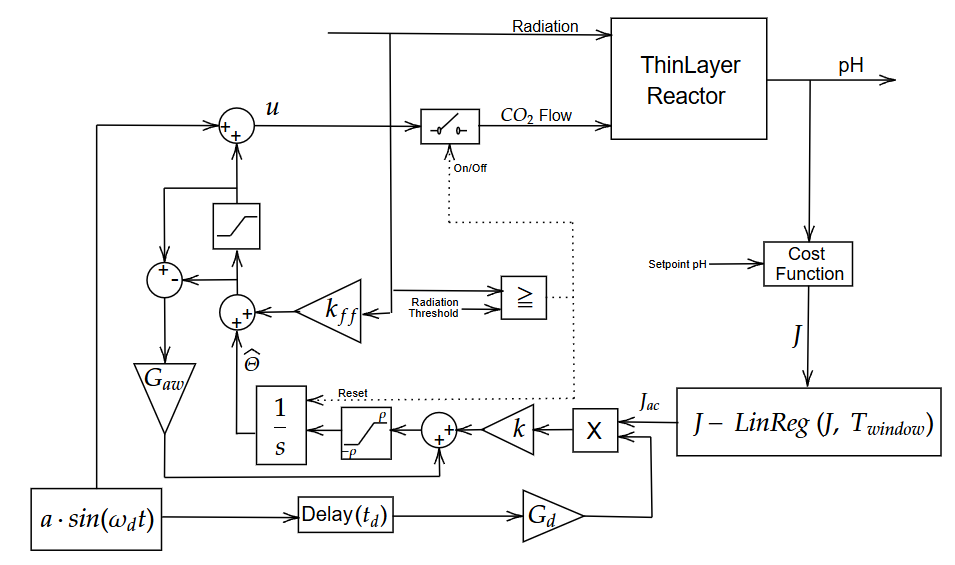}
    \caption{Block diagram of the implemented extremum seeking control (ESC) scheme. 
The conventional high-pass filtering stage is replaced by a detrending block based on moving-window linear regression. The implemented controller also includes an irradiance-based feedforward term, a saturation block to constrain 
the CO$_2$ flow-rate command, and a reset mechanism that reinitialises the integrator and
detrending window when the controller is activated.}
    \label{fig:esc_scheme}
\end{figure}

\subsubsection{Detrending-based modification}

To address these limitations, a detrending-based approach was introduced to replace the conventional high-pass filtering stage. Instead of relying on frequency-domain separation, the proposed method explicitly removes the slow-varying trend from the measured pH signal, isolating the oscillatory component induced by the dither.

The overall structure of the implemented ESC scheme is shown in Fig.~\ref{fig:esc_scheme}. 
In contrast to the classical ESC structure presented in Fig.~\ref{fig:esc_blockdiagram}, 
the implemented controller includes several practical modifications required for 
outdoor operation in the thin-layer photobioreactor.

First, the conventional high-pass filtering stage is replaced by a detrending block 
based on moving-window linear regression. This block estimates the local slow trend 
of the measured cost function and subtracts it from the raw signal, yielding the 
oscillatory component used for demodulation. The regression window was set equal to one dither period, allowing the slow trend of the pH signal to be removed while preserving the oscillatory component generated by the injected perturbation. This configuration was found to provide a reliable compromise between gradient estimation accuracy and robustness under non-stationary outdoor conditions.

Second, an irradiance-based feedforward contribution is included to account for the 
direct effect of solar radiation on photosynthetic activity and, consequently, on 
the CO$_2$ demand of the culture. This term provides a baseline CO$_2$ flow-rate 
contribution associated with the current irradiance level, while the ESC loop 
adjusts the operating point around this value based on the measured pH response.

Third, a saturation block is included at the controller output to ensure that the 
computed CO$_2$ flow-rate command remains within the admissible operating range of 
the proportional valve. This prevents unrealistic control actions and guarantees 
safe operation of the gas injection system.

Finally, a reset signal is used when the controller is activated to reinitialise both the detrending window and the ESC integrator. This avoids using outdated samples collected under different operating conditions and prevents the controller from starting with a previously accumulated adaptation state.

This approach enables a clearer separation of time scales: the slow biological dynamics are captured by the estimated trend, while the fast variations associated with the excitation signal are preserved for gradient estimation. As a result, the demodulation process becomes more robust to drift and non-stationarity.

\begin{figure*}[t]
    \centering
    \includegraphics[width=0.9\linewidth]{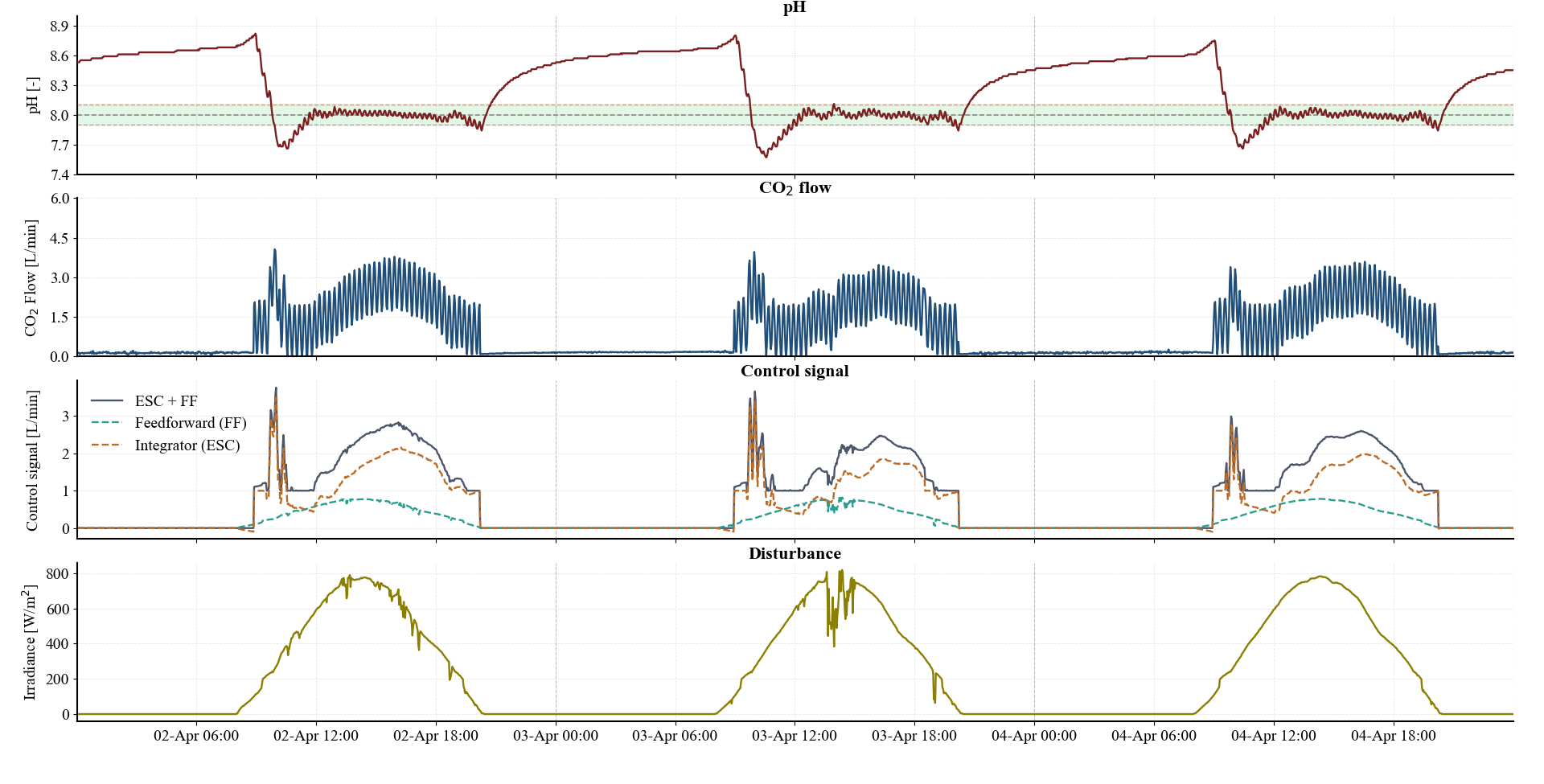}
    \caption{Closed-loop performance of the ESC with detrending-based filtering (no reset) over three days.}
    \label{fig:3days_no_reset}
\end{figure*}

The detrending operation is computationally simple and can be implemented in real time, making it suitable for integration within industrial control platforms such as PLC-based systems.

\subsubsection{Implications for gradient estimation}

The proposed modification directly improves the quality of the gradient estimation by enhancing the signal-to-noise ratio of the oscillatory component used in the ESC scheme. By reducing the influence of slow disturbances, the estimated gradient becomes more consistent and less sensitive to changes in operating conditions.

This results in improved convergence behavior and increased robustness of the ESC algorithm under realistic operating conditions, where strict stationarity cannot be assumed.






\subsection{Experimental validation}

The proposed control strategy was evaluated over three consecutive days of operation under varying irradiance conditions (see Fig.~\ref{fig:3days_no_reset}). The results demonstrate that the controller is able to maintain the pH close to the desired setpoint throughout the day--night cycles, despite the presence of strong disturbances induced by solar radiation variability.

The CO$_2$ injection profile shows the expected oscillatory behavior introduced by the ESC dither, superimposed on a slowly varying operating point.

Despite the overall satisfactory performance, a transient overshoot is observed at the beginning of each activation period due to improper initialization of the detrending filter.

\subsubsection{Filter reset strategy}

To mitigate this issue, the ESC algorithm was further modified to include a reset mechanism for the detrending filter at the moment of controller activation. By reinitializing the regression window, the trend estimation becomes immediately consistent with the current operating conditions, preventing the initial bias in the gradient estimation.

\subsubsection{Performance with filter reset}

\begin{figure*}
    \centering
    \includegraphics[width=0.9\linewidth]{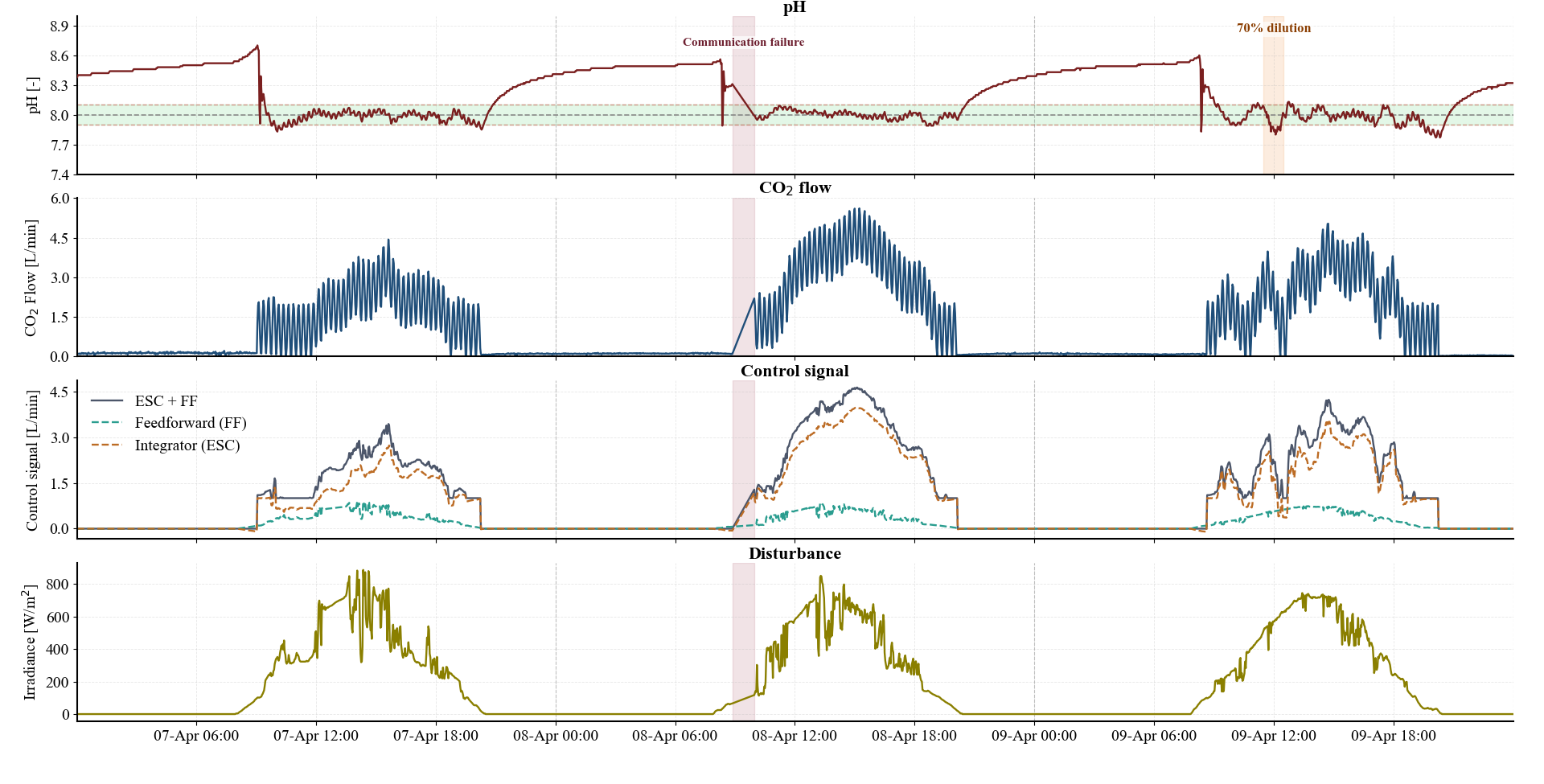}
    \caption{Closed-loop performance of the ESC with detrending and filter reset over three days.}
    \label{fig:3days_reset}
\end{figure*}

The performance of the ESC scheme with the proposed filter reset strategy is shown in Fig.~\ref{fig:3days_reset}.

The reset mechanism eliminates the initial overshoot and improves transient response. The pH remains tightly regulated around the setpoint, with small oscillations associated with the ESC excitation.

The controller adapts smoothly to irradiance variations and remains stable under disturbances, including communication failures and a 70\% dilution event, after which the system successfully re-converges.

\subsection{Performance metrics and comparison with on-off control}
\label{sec:esc_performance}

To quantitatively assess the benefits of the proposed ESC strategy, two independent three-day experiments were conducted under real conditions in thin-layer photobioreactor. In the first experiment, a conventional on-off controller was used as the baseline strategy: CO$_2$ injection was activated whenever the measured pH exceeded the setpoint $\mathrm{pH}_{\mathrm{sp}} = 8.0$, and stopped once the setpoint was recovered. In the second experiment, the proposed ESC strategy was evaluated under comparable irradiance conditions and similar biomass concentrations. The initial biomass concentrations for the ESC experiment were approximately 1.48, 1.46, and 1.45~g~L$^{-1}$ on days~1, 2, and~3, respectively, while the on-off experiment started at 1.82, 1.62, and 1.40~g~L$^{-1}$.

The performance evaluation was restricted to the periods in which the control action was active. These active intervals were determined through an irradiance hysteresis mechanism: the controller was enabled when the solar irradiance exceeded $I_{\mathrm{on}} = 100$~W/m$^2$ and disabled when it dropped below $I_{\mathrm{off}} = 20$~W/m$^2$. Let $\mathcal{T}_{\mathrm{act}}$ denote the set of time instants for which the controller remained active.

The following performance metrics were evaluated over $\mathcal{T}_{\mathrm{act}}$:

\begin{itemize}
    \item \textbf{Integral Absolute Error (IAE)}:
    \begin{equation}
        \mathrm{IAE} =
        \int_{\mathcal{T}_{\mathrm{act}}}
        \left|
        \mathrm{pH}(t)-\mathrm{pH}_{\mathrm{sp}}
        \right|dt
    \end{equation}

    \item \textbf{Total CO$_2$ consumption}:
    \begin{equation}
        \mathrm{CO}_{2,\mathrm{act}} =
        \int_{\mathcal{T}_{\mathrm{act}}}
        q_{\mathrm{CO}_2}(t)\,dt
    \end{equation}

    \item \textbf{Biomass-normalised CO$_2$ consumption}:
    \begin{equation}
        \eta_{\mathrm{bio}} =
        \frac{
        \displaystyle
        \int_{\mathcal{T}_{\mathrm{act}}}
        q_{\mathrm{CO}_2}(t)\,dt
        }{
        X_{\mathrm{avg}}
        }
    \end{equation}
    where $X_{\mathrm{avg}}$ denotes the average biomass concentration during the evaluated day.

    \item \textbf{Irradiance-normalised CO$_2$ consumption}:
    \begin{equation}
        \eta_{\mathrm{CO}_2} =
        \frac{
        \displaystyle
        \int_{\mathcal{T}_{\mathrm{act}}}
        q_{\mathrm{CO}_2}(t)\,dt
        }{
        \displaystyle
        \int_{\mathcal{T}_{\mathrm{act}}}
        I(t)\,dt
        }
    \end{equation}
\end{itemize}

The IAE metric provides a global measure of pH tracking performance by integrating the absolute deviation from the setpoint over the active control window. In addition, the biomass- and irradiance-normalised indicators compensate for differences in culture density and solar radiation between experiments, enabling a fairer comparison of carbon utilisation efficiency.

The relative improvement of the ESC strategy with respect to the on-off baseline was computed as:
\begin{equation}
    \Delta[\%] =
    \frac{
    x_{\mathrm{ESC}} - x_{\mathrm{on\text{-}off}}
    }{
    \left|x_{\mathrm{on\text{-}off}}\right|
    }
    \times 100
\end{equation}
where negative values indicate lower metric values achieved by the ESC strategy.

\begin{figure*}[h]
    \centering
    \includegraphics[width=\textwidth]{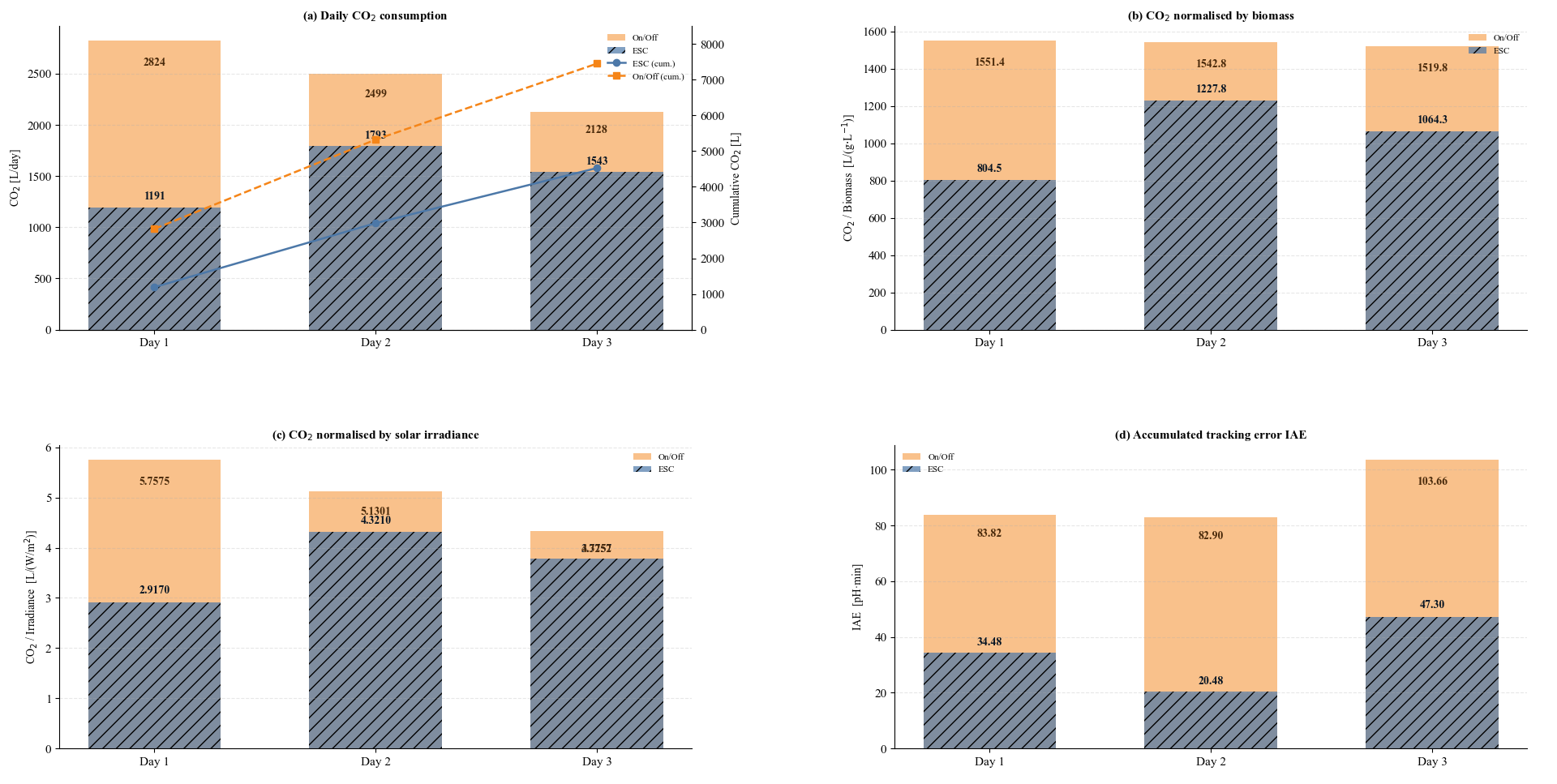}
    \caption{Comparison between the proposed ESC strategy and the conventional on-off controller over three experimental days: (a) daily and cumulative CO$_2$ consumption, (b) CO$_2$ consumption normalised by biomass concentration, (c) CO$_2$ consumption normalised by solar irradiance, and (d) accumulated pH tracking error (IAE).}
    \label{fig:esc_metrics}
\end{figure*}

Fig.~\ref{fig:esc_metrics} summarises the obtained results. The ESC consistently reduced the accumulated tracking error compared to the on-off controller, particularly during days~1 and~2, indicating tighter regulation around the desired pH setpoint. Although the improvement was smaller on day~3, the ESC still maintained a lower IAE.

Regarding carbon consumption, the ESC substantially reduced the daily CO$_2$ demand on day~1, while maintaining comparable values during days~2 and~3. The cumulative consumption trend shown in Fig.~\ref{fig:esc_metrics}(a) also indicates a lower accumulated CO$_2$ usage for the ESC throughout the experimental campaign.

The biomass-normalised and irradiance-normalised indicators further support these results. In particular, the ESC achieved significantly lower CO$_2$/biomass ratios on days~1 and~3, revealing a more efficient use of injected carbon relative to biomass concentration. Similarly, the irradiance-normalised metric confirms that the reduction in CO$_2$ consumption cannot be attributed solely to differences in solar radiation conditions.

In summary, the proposed ESC strategy achieved improved pH regulation together with a more efficient utilisation of CO$_2$, which is particularly relevant since CO$_2$ supply represents one of the major operating costs in microalgae production. These results demonstrate the suitability of the proposed approach for outdoor operation in thin-layer photobioreactors.

\begin{table*}[htbp]
\centering
\small
\begin{tabular}{lcccccccccccc}
\hline
& \multicolumn{3}{c}{CO$_2$ [L]} 
& \multicolumn{3}{c}{CO$_2$/Biomass [L\,g$^{-1}$]}
& \multicolumn{3}{c}{$\eta_{\mathrm{CO}_2}$ [L/(W/m$^2$)]}
& \multicolumn{3}{c}{IAE [pH$\cdot$min]} \\
\cmidrule(lr){2-4}
\cmidrule(lr){5-7}
\cmidrule(lr){8-10}
\cmidrule(lr){11-13}
Day 
& ESC & On/Off & $\Delta[\%]$
& ESC & On/Off & $\Delta[\%]$
& ESC & On/Off & $\Delta[\%]$
& ESC & On/Off & $\Delta[\%]$ \\
\hline
1 
& 1191 & 2824 & -57.8
& 804.5 & 1551.4 & -48.1
& 2.917 & 5.758 & -49.3
& 34.48 & 83.82 & -58.9 \\
2 
& 1793 & 2499 & -28.3
& 1227.8 & 1542.8 & -20.4
& 4.321 & 5.130 & -15.8
& 20.48 & 82.90 & -75.3 \\
3 
& 1543 & 2128 & -27.5
& 1064.3 & 1519.8 & -30.0
& 3.776 & 4.353 & -13.3
& 47.30 & 103.66 & -54.4 \\
\hline
Total / Mean
& 4527 & 7451 & -39.2
& 1032.2 & 1538.0 & -32.9
& 3.671 & 5.080 & -27.7
& 34.09 & 90.13 & -62.2 \\
\hline
\end{tabular}
\caption{
Performance comparison between the proposed ESC strategy and the conventional on-off controller over three experimental days. Metrics were computed only during the active control periods defined by the irradiance hysteresis. Negative values of $\Delta[\%]$ indicate an improvement of the ESC strategy with respect to the conventional on-off controller.
}
\label{tab:esc_performance}
\end{table*}

\section{Conclusions}

This work presented the experimental implementation of a model-free extremum seeking control (ESC) strategy for pH regulation in an thin-layer photobioreactor. Unlike conventional approaches based on fixed switching logic or explicit process models, the proposed controller relies exclusively on real-time measurements to adapt the CO$_2$ injection rate under strongly time-varying operating conditions.

The experimental characterization of the reactor highlighted the significantly faster dynamics of Thinlayer systems compared to conventional raceway reactors, enabling the use of higher ESC excitation frequencies while preserving reliable gradient estimation. However, these fast dynamics also revealed important limitations of standard ESC filtering approaches under non-stationary conditions.

To address this issue, a detrending-based signal conditioning stage was introduced, replacing the conventional high-pass filter. In addition, a reset mechanism for the detrending window was incorporated to improve transient behavior during controller activation. The resulting architecture demonstrated robust operation under varying irradiance conditions, communication interruptions, and large biomass dilution events.

The comparative experimental evaluation against a conventional on-off controller showed that the proposed ESC strategy significantly improves both regulation performance and carbon utilization efficiency. In particular, the ESC reduced cumulative CO$_2$ consumption by approximately 39\% while decreasing the accumulated pH tracking error by more than 60\% over the evaluated operating periods. Biomass and irradiance normalized indicators further confirmed that these improvements were not solely attributable to differences in environmental conditions.

Overall, the obtained results demonstrate that model-free ESC constitutes a promising alternative for advanced control of fast and highly nonlinear photobioreactor systems, where explicit modeling is difficult and operating conditions continuously evolve.

Future work will investigate the application of the proposed ESC framework to dissolved oxygen (DO) regulation, as well as the development of adaptive gain strategies to accelerate convergence under varying irradiance conditions. In particular, adapting the ESC tuning parameters according to solar radiation levels could improve transient response and optimization performance during rapidly changing environmental conditions.

\section*{Acknowledgments}
This work has been financed by the following projects: PID2023-150739OB-I00 and PDC2025-165379-I00 financed by the Spanish Ministry of Science and also by the European Union (Grant agreement IDs: 101060991, REALM; 101214199, ALLIANCE).

\bibliographystyle{elsarticle-harv}
\biboptions{authoryear}
\bibliography{references}
\end{document}